# Automated Exploration of Reaction Network and Mechanism via Meta-dynamics Nanoreactor

*Yutai Zhang[1], Chao Xu[1], Zhenggang Lan[1,\*]*

[1] Guangdong Provincial Key Laboratory of Chemical Pollution and Environmental Safety and MOE Key Laboratory of Environmental Theoretical Chemistry, SCNU Environmental Research Institute, School of Environment, South China Normal University, Guangzhou 510006, P. R. China.

**KEYWORDS**: meta-dynamics, reaction mechanism, reaction network, nanoreactor

# Abstract

We developed an automated approach to construct the complex reaction network and explore the reaction mechanism for several reactant molecules. The nanoreactor type molecular dynamics was employed to generate possible chemical reactions, in which the meta-dynamics was taken to overcome reaction barriers and the semi-empirical GFN2-xTB method was used to reduce computational cost. The identification of reaction events from trajectories was conducted by using the hidden Markov model based on the evolution of the molecular connectivity. This provided the starting points for the further transition state searches at the more accurate electronic structure levels to obtain the reaction mechanism. Then the whole reaction network with multiply pathways was obtained. The feasibility and efficiency of this automated construction of the reaction network was examined by two examples. The first reaction under study was the $HCHO + NH_3$ biomolecular reaction. The second example focused on the reaction network for a multi-species system composed of dozens of HCN and $H_2O$ compounds. The result indicated that the proposed approach was a valuable and effective tool for the automated exploration of reaction networks.

# 1. INTRODUCTION

The exploration and investigation of complex and unknown chemical reaction mechanisms is one of main goals in various fields.[1-3] Over decades, the tasks of the reaction mechanism explorations were often achieved manually by erudite researchers using their chemical knowledge and intuition. This approach may not be efficient, because it highly relies on extensive manual calculations. Nowadays, with the rapid progresses of the computer hardware and the application of novel theoretical methods, the exhaustive exploration of reaction space becomes feasible. Various automated computational strategies have been developed to explore the reaction mechanisms effectively.[4-6]

In theoretical studies of reaction mechanisms, one central task is the search of transition state (**TS**) structures. **TS** is the saddle point in the minimum-energy pathway (MEP) that connects reactants and products. The information of the **TS** is the starting points for the understanding of the reaction mechanisms and the link to experimental studies. Except traditional manual **TS** search starting from the given initial guess structure, some automated **TS** searching approaches were also proposed.[7-16] When the reactions that reactants and products are known, it is accessible to locate the **TS** structure by applying the double-end searching approaches such as the nudged elastic band (NEB) method,[7-10] the growing string method (GSM),[11, 12] the freezing string method (FSM)[13] and other string methods.[14-16] These methods can offer a path approximate to MEP and initial guess for **TS** searching task.

In practical terms, the investigation of possible reaction pathways with unknown products is challenging and essential, which motivates the development of various novel algorithms and methods. Starting from given reactants, it is possible to explore reaction by using knowledge-based approaches to generate reaction templates such as NetGen,[17] RMG (reaction mechanism generator),[18, 19] Kinbot,[20] or focusing on the information of potential energy surface (PES) to search **TS** such as GRRM (global reaction route mapping) strategy[21] with ADDF (anharmonic downward distortion following) method[22, 23] and AFIR (artificial force induced reaction) method[24-27] and SSW (stochastic surface walking) method.[28-30] In addition to these methods, various approaches[31-36] that generate plausible intermediates and products starting from the given reactants were developed to explore the reaction space and predict the reaction mechanisms. Some theoretical works employed the molecular connectivity graph to generate possible products, which includes ZStruct method,[37, 38] ZStruct2 method,[39] ReNeGate,[40] YARP (yet another reaction program)[41, 42] and other methods.[43-45]

Notably, the exploration of the complex reaction space may be realized by the reactive molecule dynamics (MD).[46-56] The reaction events in the reactive MD clearly demonstrate the chemical transformations from reactants to products, which can be used as the initial guesses for **TS** searching tasks. However, one of challenges in the reactive MD approach is to produce enough reaction events with the manageable computational efforts. Therefore, two challenging aspects of the problem should be considered.

First, it is necessary to increase the frequency of reaction events. For instance, this purpose is realized by introducing high temperature,[48, 49] high pressure[50] and both. The famous nanoreactor[51-56] method that combines both was proved to be highly effective. Additionally, it is possible to introduce external potential/force[57-62] to accelerate the emergence of reaction events, in which one practical approach is applying the biasing potential. In the meta-dynamics (MTD),[63-65] the biasing potential is added along the collective variables (CVs), which can be understood as reaction coordinates. In this case, the trajectory can escape from one local energy minima to reach others, giving possible reaction products. However, the CV selection is not a trivial task, as it deeply modifies how the new configurations will be generated. Various possible ways were proposed to define CVs, while more or less required priori chemical knowledge.[66]

Second, it is important to achieve the balance between computational cost and accuracy in reactive MD approaches. The employment of the ab initio quantum mechanics (QM) methods in the MD is certainly highly preferable due to their accuracy. However, this results in large computational cost. Even when the Hartree−Fock (HF) theory is taken, the computational cost is still rather high, when the system becomes larger. The MD based on empirical force fields like Reactive Force Field (ReaxFF)[46, 47, 49, 54, 67-69] can achieve the high efficiency in the MD simulation, while the accuracy of this approach is not comparable with the one at the quantum chemistry level. One possible solution for the above two problems is to employ the fragment-based reactive explicit polarization (ReX-Pol) approach in the direct molecular dynamics.[70]

Alternatively, the density-functional based tight-binding (DFTB) method[71-75] may provide a preferable balance on both issues.

Recently, Grimme[76] proposed a MTD method combining a DFTB3 method[73-75] variant termed GFN-xTB (geometry, frequency, non-covalent, extended tight binding) method[77] and improved GFN2-xTB method,[78] which alleviated above problems in the reactive MD method. This MTD method uses the root-mean-square deviation (RMSD) in the Cartesian space to define the CVs. In this way we do not need to know the CV in advance, and it is suitable to discover unknown reactions. Additionally, with the highly efficient GFN2-xTB method, the long-time and large-scale simulation of the rather large system with hundreds of atoms can be performed on the ground-state PES in the qualitative or even semi-quantitative manner. It was proven to give the proper descriptions on various chemical reactions, such as thermal decomposition, oligomerization, oxidation, thermally forbidden dimerization, and multistep intramolecular cyclization reaction, and a Miller-Urey model system.[76]

To understand the detailed mechanism of the explored reaction space in reactive MD simulations, it is necessary to analyze reaction events. However, due to the long-time propagation and large system scale in the reactive MD simulations, it is not practical to only reply on manual efforts to extract the numerous reaction events and to identify the relevant chemical species from reactive trajectories. Therefore, some efficient trajectory analysis methods were developed, which rely on the time-series analysis of bond order[55, 79] or molecular connectivity[46, 57, 80] to identify the reaction events.

Inspired by the aforementioned studies, we developed an automated approach to explore reaction network. First, we employed MTD simulation to produce reactive trajectories containing ample reaction events to explore the reaction space exhaustively. Here the MTD was run at the GFN2-xTB level, which highly reduced computational cost. Second, we examined the molecular connectivity graphs in reactive trajectories to identify molecules participated in the reaction events and obtain the occurrence of reaction events with the hidden Markov model (HMM).[46, 57, 80, 81] Third, based on the information of molecular connectivity, we tried to identify the reactants and products in each reaction events by iteratively searching relevant fragments and involved atoms. Fourth, we applied the **TS** location strategy to explore the reaction mechanisms. The vibrational analyses of **TS** structures and intrinsic reaction coordinate (IRC) calculations helped to screen the proper **TS** structures and construct the corresponding MEPs. Finally, two different reaction networks were built. The first one was the chemical transformation network composed of reaction events found in the MTD simulations, and the second one focused on the reaction mechanism network based on discovered MEPs.

We applied the developed approach to investigate two typical reactions. The first was the simple bimolecular reactions between formaldehyde (HCHO) and ammonia ($NH_3$) molecules, which were investigated previously by Zimmerman[37] and ourselves.[57] We selected this system as a prototype to examine the effectiveness of our approach.

The second was the reactions between dozens of hydrogen cyanide (HCN) and water (H$_2$O) molecules. As an important species in prebiotic earth, HCN can initialize a complicated reaction network with many synthetic routes, which is also thought to play a critical role in the origin of life. Considerable efforts were devoted to study the reactions of this species, including the HCN polymerization reaction[31, 55] and the bimolecular reactions[82, 83] between HCN and H$_2$O. Vanka et.al[84] used the ab initio high-temperature and high-pressure nanoreactor method to study the system composed of dozens of HCN and H$_2$O compounds. Therefore, we tried to study the reaction network of this system. The studies on both reaction systems indicated that our current approach was a powerful approach to discovery the complex reaction network automatically, which also achieved the excellent balance between computational cost and reliability.

## 2. METHODS

We proposed an automated approach to construct chemical reaction networks and to explore mechanisms. Starting from reactant species, the MTD simulation step was designed to explore the reaction space, which provided the reactive trajectories with numerous reaction events. Next, the reaction events were identified and extracted from these trajectories, which provided the important initial guess for the **TS** location and MEP search. Last, based on reaction events and MEPs, two different types of reaction networks were constructed. The workflow of our approach is shown in Figure 1 and the details of above steps are discussed below.

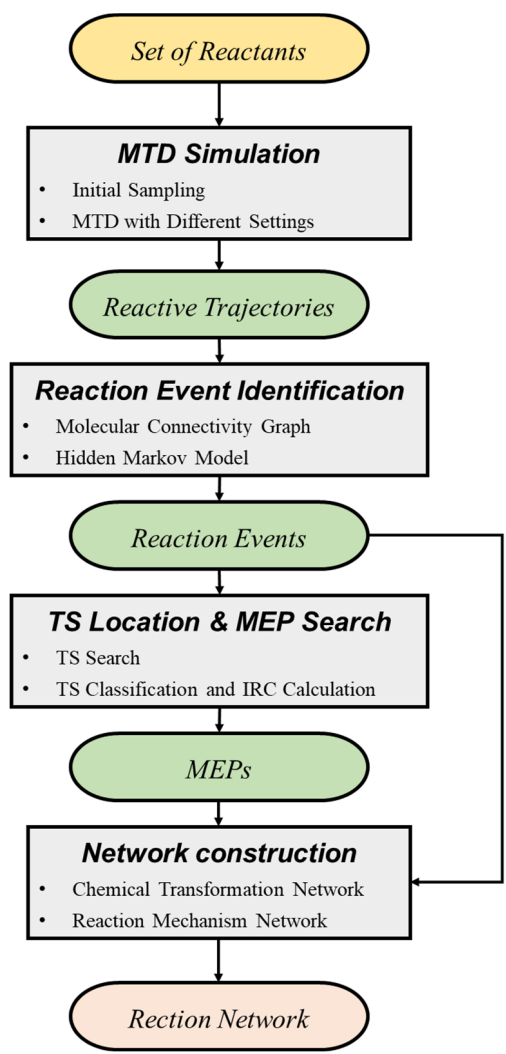

**Figure 1.** The workflow diagram in this work.

## 2.1. MTD Simulation at GFN2-xTB level

In this approach, we first employed the MTD simulation at the GFN2-xTB level to explore the reaction space. This step included the initial sampling and the MTD simulation with different settings.

### 2.1.1 Initial Samplings

In this step, we created the proper reactive system as a reactor and performed the suitable initial samplings. In this work, a spherical system with the boundary was defined, which contains all reactant molecules. We defined the composition of reactant molecules and the density of the entire system. According to these information, a spherical system composed of given reactant molecules was built by using our homemade Python script linked with the Packmol software package.[85] After the packing configuration was constructed, we ran the geometry optimization of the whole system at the GFN2-xTB level.

Based on the lowest-energy structure, we conducted a long-time MD simulation at the GFN2-xTB level to provide the initial sampling configurations. To ensure the molecules keeping inside of the reactor, the MD simulation introduced a wall potential whose radius was based on the sphere radius mentioned above. The details of the wall potential are given in the Supporting Information. The SHAKE algorithm[86] was applied to avoid the breaking of the covalent bonds. The whole MD part was divided into two successive steps, i.e., equilibrium run and product run. We first monitored the temperature of the system to make sure that the whole system reaches to equilibrium, and then the MD with the constant temperature was employed as initial sampling configurations.

### 2.1.2. MTD Simulations

In this work, we applied the MTD (RMSD) / GFN2-xTB method proposed by Grimme.[76] The wall potential was applied to construct the whole spherical system as a reactor. The application of the suitable biasing potential was also essential in the MTD simulation. Following the important advice in previous studies by Grimme,[76, 87] we tried to determine the proper settings in the MTD simulation by the trial-and-test calculations. More detailed discussions on the MTD simulation are provided in the Computational Details and the Supporting Information.

**2.2. Reaction Event Identification**

To analyze the MTD trajectories containing many reaction events, we employed an automated reaction event identification method, inspired by previous works,[46, 57, 80] First, we got the atomic connectivity information during the trajectory propagation to construct the state sequences of molecular graphs. Next, we applied the two-state HMM to filter the "noise" in the state sequences to obtain the occurrence time of reaction events. Finally, we identified all the reactants and products in each reaction event. More details are discussed as follows.

**2.2.1. Molecular Connectivity Graph**

At different time in the trajectory propagation, the structure of the entire system was analyzed by the examination of the atomic connectivity obtained from the pair-wise atomic distances. Here we used the Open Babal[88] Python module to judge the connectivity. In Open Babel, the formation or the break of a chemical bond between a pair of atoms ($i, j$) with the atomic distance $r_{i,j}$ is judged as existence according to the

below relation:

$$0.4 \text{ Å} < r_{i,j} < R_i + R_j + 0.45 \text{ Å}$$

where $R_i$ and $R_j$ are the default atomic covalent radius of the atom *i* and atom *j*, respectively. After the acquisition of connectivity information, we converted it into a connectivity graph by employing NetworkX[89] Python module. In principle, a connectivity graph may contain various subgraphs and each subgraph should represent a molecular fragment. The nodes and edges of the subgraph represents the atoms and the bonds, respectively. For each individual subgraph, we detected whether it appeared or vanished along the trajectory propagation and such information defined a time-dependent state sequence composed of many 0 and 1 numbers. The signal 0 at the specific time *t* means that a particular subgraph is not detected at present time, while the signal 1 indicates the reversed information. The 1→0 signal transition means that the corresponding fragment vanishes within two successive time steps, which indicates that it should participate in a reaction event as a reactant. Similarly, the signal transition of 0→1 means that the fragment forms as a product. The state sequence of each subgraph were processed in next step.

### 2.2.2. HMM

The biasing potential in the MTD might cause the high kinetic energy and fast molecular vibrations. As a result, the state sequence might show the fast oscillation between 0 and 1, which could be viewed as "noise". If the "noise" is not filtered, a large number of state changes have to be assigned as reaction events, which significantly increases the workload in the step of the **TS** location. Here we applied a two-state HMM to filter those "noise".

The HMM is based on the Markov chain. In a time-dependent state sequence $\{X_0, X_1, ..., X_{t-1}, X_t\}$, the current state $X_t$ is only determined by the previous state $X_{t-1}$. The probability of moving from state $X_{t-1}$ to state $X_t$ is defined by the transition probability $T = P(X_t | X_{t-1})$. In HMM, the state $X_t$ cannot be observed directly, which is called the hidden state. A hidden state $X_t$ can generate an observable state $V_t$ according to the output probability $O = P(V_t | X_t)$, giving a sequence of the observable states $\{V_0, V_1, ..., V_t\}$.

For the two-state HMM, both hidden and observable states only include two states. An initial probability distribution $\pi$ $\{\pi_1, \pi_2\}$ represents the possibility that the hidden state $X_0$ starts in *State 1* or *State 2*. The transition probability $T$ and the output probability $O$ can be represented by a 2 by 2 square matrices, which are called the transition matrix and the output probability matrix, respectively.

In this work, the state sequence got in the previous step is the sequence of the observable states. Our goal is to find the most-likely hidden state sequence from the observable state sequence by using the Viterbi algorithm.[90] In this work, the below parameters were employed in the HMM according to the recommendation of previous works:[46, 80]

$$\begin{cases} \pi = \{0.5, 0.5\} \\ T = \begin{pmatrix} 0.999 & 0.001 \\ 0.001 & 0.999 \end{pmatrix} \\ O = \begin{pmatrix} 0.6 & 0.4 \\ 0.4 & 0.6 \end{pmatrix} \end{cases}$$

Based on the HMM to analyze the state sequence, we might obtain the occurrence time of reaction events, and then identify all reactants and products to construct the elementary reaction events.

**2.2.3. Identification of Reactants and Products for a Reaction Event**

Our next task was to identify all reactants and products for each reaction event. We briefly discussed the scheme here, and more details are given in the Supporting Information. Starting from the time in which a reaction event takes place, we traced all involved atoms over 30 MTD dump steps backwards along the time axis to obtain the molecular connective graph before the reaction event, and then traced forwards 30 steps to get the connective graph after the reaction events. The iterative process of comparing connective graphs was employed to define which atoms were directly involved in the reaction events. Finally, after acquiring all involved atoms, we transformed these involved atoms and their connectivity information into canonical SMILES (simplified molecular input line entry specification)[91] by employing the RDKit[92] Python module. Since the SMILES codes were character strings, each elementary reaction event was represented by a string.

**2.3. TS Location and MEP Search**

In this step, we re-examined each reaction event again to obtain the refined reaction pathway by searching their **TS**s and MEPs. Here, a practical strategy was employed to locate the **TS**s.

In the **TS** location strategy, we selected the guess structures composed of involved atoms directly from the MTD snapshots. Totally seven frames centered at the reaction time were chosen with the time interval of the 10 MTD dump time steps, as shown in Figure 2. All seven structures were chosen as initial guesses to perform the **TS** optimization.

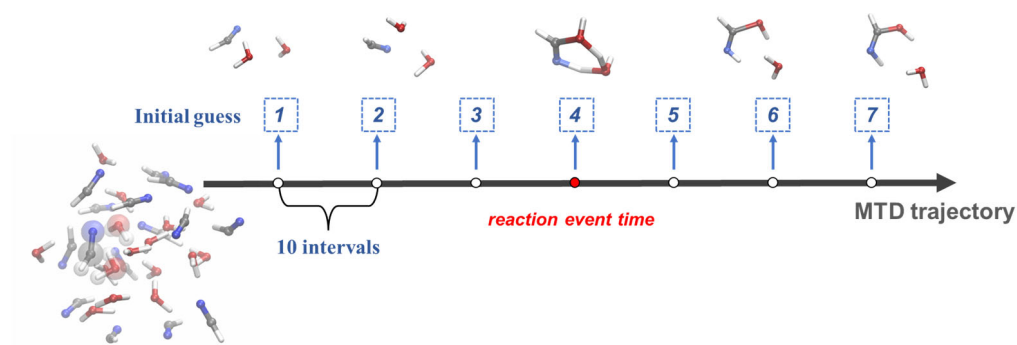

**Figure 2.** The selection of the initial guesses for the transition state (**TS**) search. For a reaction event, we select a geometry every 10-time intervals and collect three geometries before the reaction time event. In addition, we also trace forwards with time being and select three geometries in the similar manner. By including the geometry at the reaction time, totally seven structures equally distributed in the time axis are extracted as the initial guesses to perform the **TS** optimization.

Because these fragments were extracted from their surrounding environment, the charges and spin multiplicities of them needed be determined. We tried to employ a practical strategy to estimate these values. For all chosen structures, the entire system was calculated by using electronic structure calculations at the GFN2-xTB level. The Mulliken charges at each atom were calculated and the sum of these values over all involved atoms defined the net charge of the corresponding fragment. After averaging the charge of the chosen fragment over seven structures and rounding such value to the nearest integer, the charge of the fragment was obtained. According to it, the lowest possible spin multiplicity was set. This meant that some high spin fragments were not considered in the current method.

To avoid the extensive **TS** search, a useful trick was employed to reduce the computational cost here. As discussed previously, each elementary reaction was represented by a string and many similar elementary reactions may appear in the MTD run. After the collections of all reactions over trajectories, we divided these reactions into several groups by the clustering analyses of the similarity of their corresponding strings. This step finally defined how many different elementary reactions were found in the trajectory propagations. For each elementary reaction, we randomly chose 10 reaction events (totally 70 structures) to perform the **TS** search task. When some reaction events appeared less than 10 times, we chose all of them in the **TS** search job. Then the direct **TS** optimizations at the more accurate electronic-structure levels were applied. If the **TS** optimization was not converged, we might employ some double-end methods like the climbing image NEB (CI-NEB) method[9, 10] to search the **TS**.

After the **TS** optimization, the frequency analyses were conducted to confirm the validation of the **TS** structures. For the same reaction event, the located **TS**s may be very similar. Here, we collected all located **TS** structures, and tried to assign them into different **TS** groups. To achieve this task, we used the agglomerative cluster approaches in Scikit-learn[93] Python module by examining their energy and vibrational frequencies. For each represented **TS** structure, IRC calculations were applied to find the proper pathways connecting to reactants and products. The reasonable IRC paths were selected as MEPs.

**2.4. Network Construction**

In this step, we considered to build two different types of reaction networks. The first one was the chemical transformation network constructed directly according to the reaction events identified in the MTD trajectories, while the second one was the reaction mechanism network built in the basis of the MEPs.

In the first type of reaction networks, we only replied on the species involved in the MTD trajectories and used their SMILES codes to build the reaction network. Here the whole network was represented by a directed graph, in which its nodes represented the reaction species, and the directed edges represented the chemical transformation between species. The network was constructed by employing the NetworkX Python module in following manners:

- For a unimolecular decomposition reaction, such as A => B + C, the nodes represent the species A, B, and C, and the directed edges represent the reaction direction for A => B and A => C.

- For a bimolecular addition reaction, such as A + B => C, the nodes representing A, B, and C are registered, and edges representing the reaction direction of A => C and B => C are registered.

- For a more complex reaction, we trace all involved atoms to determine the reaction direction. For instance, we may monitor all atoms of the reactant A. If they appear in products D and E, the edges of A => D and A => E are registered.

For all nodes, their sizes and colors were decided by the number of times of the appearance of corresponding species in reaction events. For all edges, their colors and widths were determined by the number of times of the corresponding reaction events. Furthermore, rare reaction events were filtered to prune the graph more concise. The Fruchterman–Reingold force-directed algorithm[94] in the NetworkX Python module was used to reduce the overlap between different nodes in the network and enhance its readability.

In the second type of reaction networks, we tried to build the reaction mechanism network based on MEPs. This network contained not only the structure information of reactants and products, but also the activation energy $E_a$ information of reactions. In this work, the activation energy was defined as the difference in electronic energy between the reactants and the corresponding **TS**. For simple systems, we might put all available reactions together and provide a whole reaction mechanism network. For the complicated systems, it was not transparent to show all involved reactions in a single picture. In this case, we tried to build the reaction mechanism network according to a specific group of reactions.

Two types of networks emphasize different information. The chemical transformation network constructed from the SMILES code of reaction events is suitable for describing the continuous process of chemical transformations in MTD simulation and can be automatically generated after the identification of reaction events without requiring high-level QM calculations. This network is ideal for understanding high-energy reactions involving many species and channels, such as combustion reactions. As the contrast, the MEP-based reaction mechanism network is relatively complex, because it requires high-cost QM calculations and additional manual judgments. However, it provides the clear information on the structures of the **TS**, the stable structures of the reactants and products, and the values of the barrier for important reactions. It is suitable for the detailed understanding of the reaction mechanism.

## 3. COMPUTATIONAL DETAILS

In this work, the geometry optimization, MD for initial sampling, MTD simulations and electronic structure calculations for Mulliken charge were performed at the GFN2-xTB[78] level of theory in the xtb software package.[95] The details of initial sampling for two systems under studying are given in the Supporting Information.

In the MTD simulations, we conducted multiple sets of test calculations with different parameter settings. Normally, we should avoid the situations with too less number of reactive MTD trajectories, and this indicates that the biasing potential is too low. Simultaneously, if the MTD test calculations become unstable in some parameter settings, the biasing potential might be too large. Test calculations provided us some judgement on how to choose suitable parameters in the actual MTD run. The details of MTD setting tests for two specific reactions studied are given in the Supporting Information.

Subsequently, we proceeded with the actual MTD simulations using the selected parameter settings. First, we ran two groups of trajectories (main group and validation group) with the same parameter setting. Typically, a significant number of reaction events were obtained from these trajectories. Some events might occur repeatedly, while others might only take place once or twice. We assumed that the reactions occurring very frequently in the MTD represented the major reactions. If the major reactions found in the trajectories of the validation group were already discovered in the main group, we believed that the calculations with this chosen parameter setting are converged. Second, we also needed to run the MTD simulations with different parameter settings to judge the convergence. In the case of the simple bimolecular reactive systems, if the major reactions events observed in the low-biasing-potential MTD calculations were also identified in high-biasing-potential MTD calculations, we assumed that the calculations were converged. For the complex systems with many reactive species, sometime the polymerization might take place and the large-sized species might yield. In such cases, we only compared the reaction events that produce the small-sized species under different biasing potential according to the above rules used for small systems. If the major events in different settings were similar, we assumed that the calculations were converged.

The above discussions provide general guidelines to perform the MTD simulations. More technical details, including the MTD treatment of two typical reactions under study, are given in the Supporting Information.

In the MEP search task, the Gaussian 16 software package[96] was used to perform all **TS** geometry optimizations, vibrational frequency calculations and IRC calculations at the B3LYP[97-99]/def2-SVP[100, 101] level of theory with the GD3BJ[102] dispersion correction.

## 4. RESULTS AND DISCUSSION

### 4.1. HCHO and $NH_3$ Reaction

As a simple test case, the bimolecular reactions between HCHO and $NH_3$ were investigated. The chemical transformation network based on major reaction events are given in the Supporting Information. We constructed the whole reaction mechanism network based on MEPs, as shown in Figure 3.

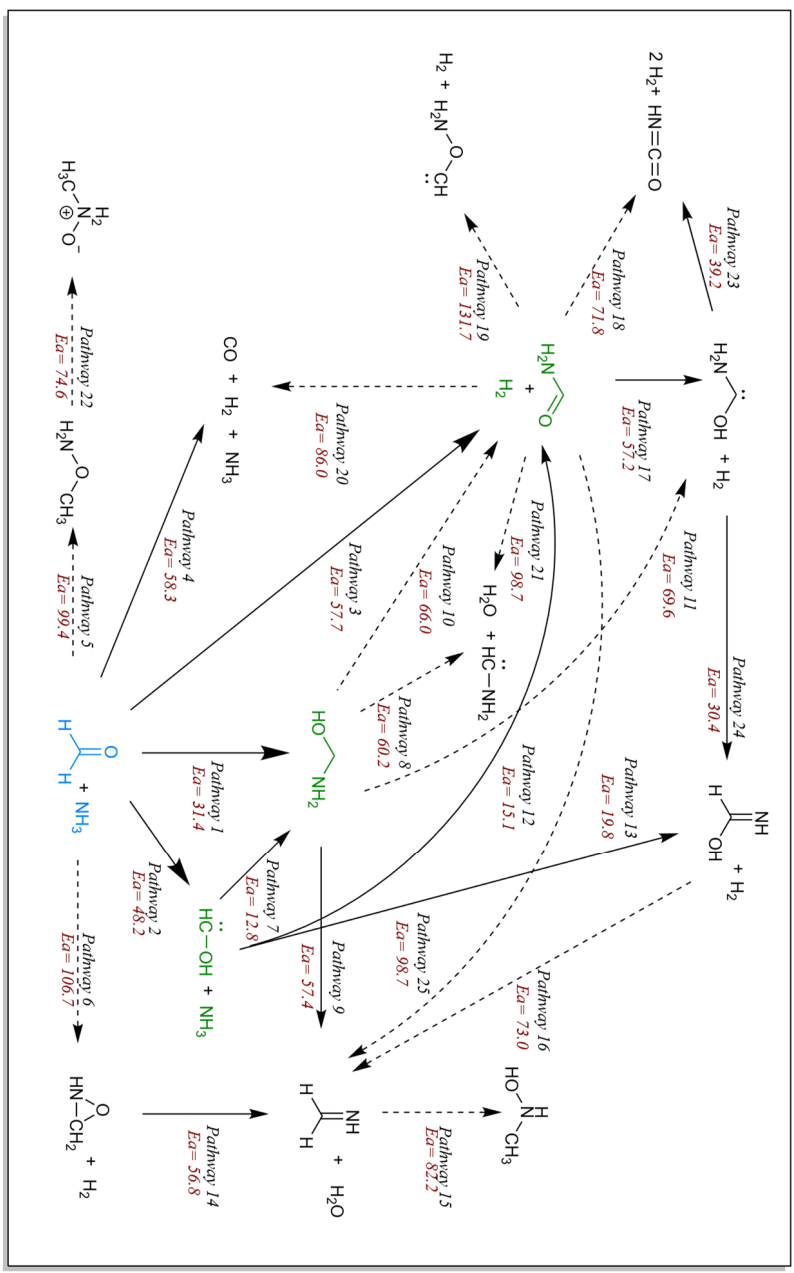

**Figure 3.** The reaction mechanism network starting from HCHO and NH$_3$ (highlighted in blue). Each elementary reaction pathway is labeled by its index number, and the corresponding reaction barrier is given in the unit of kcal/mol. The dashed arrow line indicates a barrier higher than 60 kcal/mol.

Starting from the reactants of HCHO and NH$_3$, many reactions were found and several of them show high barriers larger than 60 kcal/mol. Next, we mainly focused on the primary reactions between HCHO and $NH_3$, totally six important elementary reactions (Pathways 1-6) were found. The **TS** structures and product structures of these six pathways are provided in the Supporting Information.

Pathway 1 is the N-H addition reaction that yields aminomethanol ($NH_2CH_2OH$). This pathway displays the lowest reaction barrier (31.4 kcal/mol) starting from HCHO and $NH_3$. The compound $NH_2CH_2OH$ should be stable, because its self-decomposition reactions show rather high barriers above 60 kcal/mol (Pathway 8, 10, 11). Pathway 2 is the HCHO isomerization reaction to give the unsaturated alcohol CHOH with the presence of a catalytic $NH_3$ molecule. CHOH is highly unstable, which can easily react with $NH_3$ to form other species via three low-barrier (< 20 kcal/mol) pathways (Pathway 7, 12, 13).

Pathway 3, 4 display the similar barrier heights in the range of 50~60 kcal/mol. Pathway 3 yields carbamoyl ($NH_2CH=O$) and $H_2$. These species are important in the network. At the same time, the further reactions of the product in Pathway 1, 2 can also yield these species. Pathway 17, 19 show that the isomerization of $NH_2CH=O$ generates unsaturated amino-alcohols and amino-alkenes, while Pathway 18, 20 show the dehydrogenation reactions of $NH_2CH=O$. Pathway 4 yields the small species $H_2$, CO and $NH_3$, and the reaction barrier of it is 58.3 kcal/mol, while the forming these species via the deamination reaction of $NH_2CH=O$ is unfavorable (Pathway 20).

Pathway 5, 6 show very high barriers larger than 90 kcal/mol. Pathway 5 is the N-

O addition pathway to generate methoxyamine ($CH_3ONH_2$) and Pathway 6 produces three-membered ring $CH_3NO$ and $H_2$.

Four pathways (Pathway 1, 2, 4, 5) were mentioned in the previous work by Zimmerman,[37] while some differences existed in detail. The lowest barrier reaction we found (Pathway 1) is consistent with the previous study of Zimmerman. While the barrier of Pathway 4 (58.3 kcal/mol) is lower than the barrier of identical reaction in the work by Zimmerman, which is 80 kcal/mol. Additionally, Pathway 3 and Pathway 6 were also reported in our previous work.[57]

**4.2. HCN and H$_2$O Reaction**

In the second example, the multi-species reactions between dozens of HCN and $H_2O$ molecules were studied. We constructed the chemical transformation networks based on the major reaction events that occurred more than twice in all MTD trajectories, as shown in Figure 4.

**(a)**

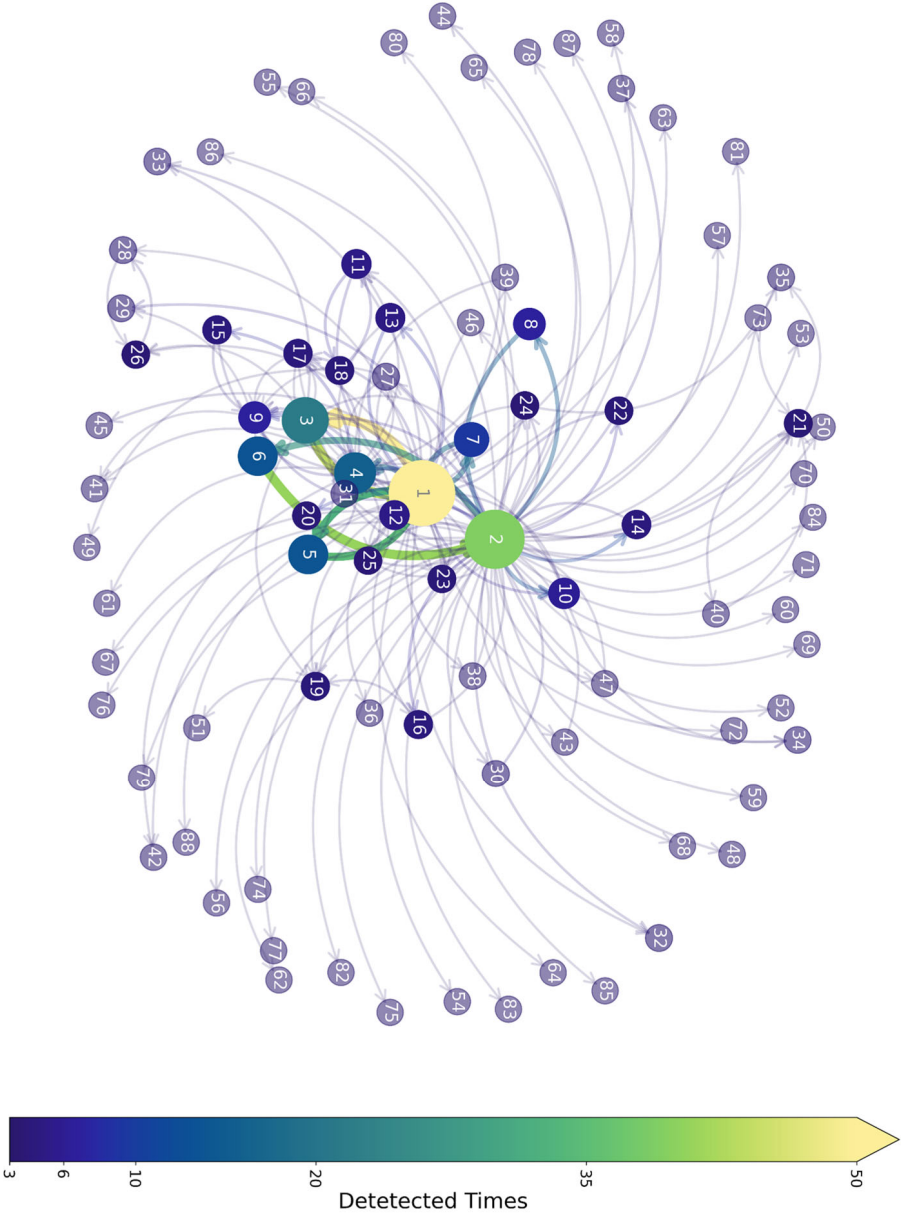

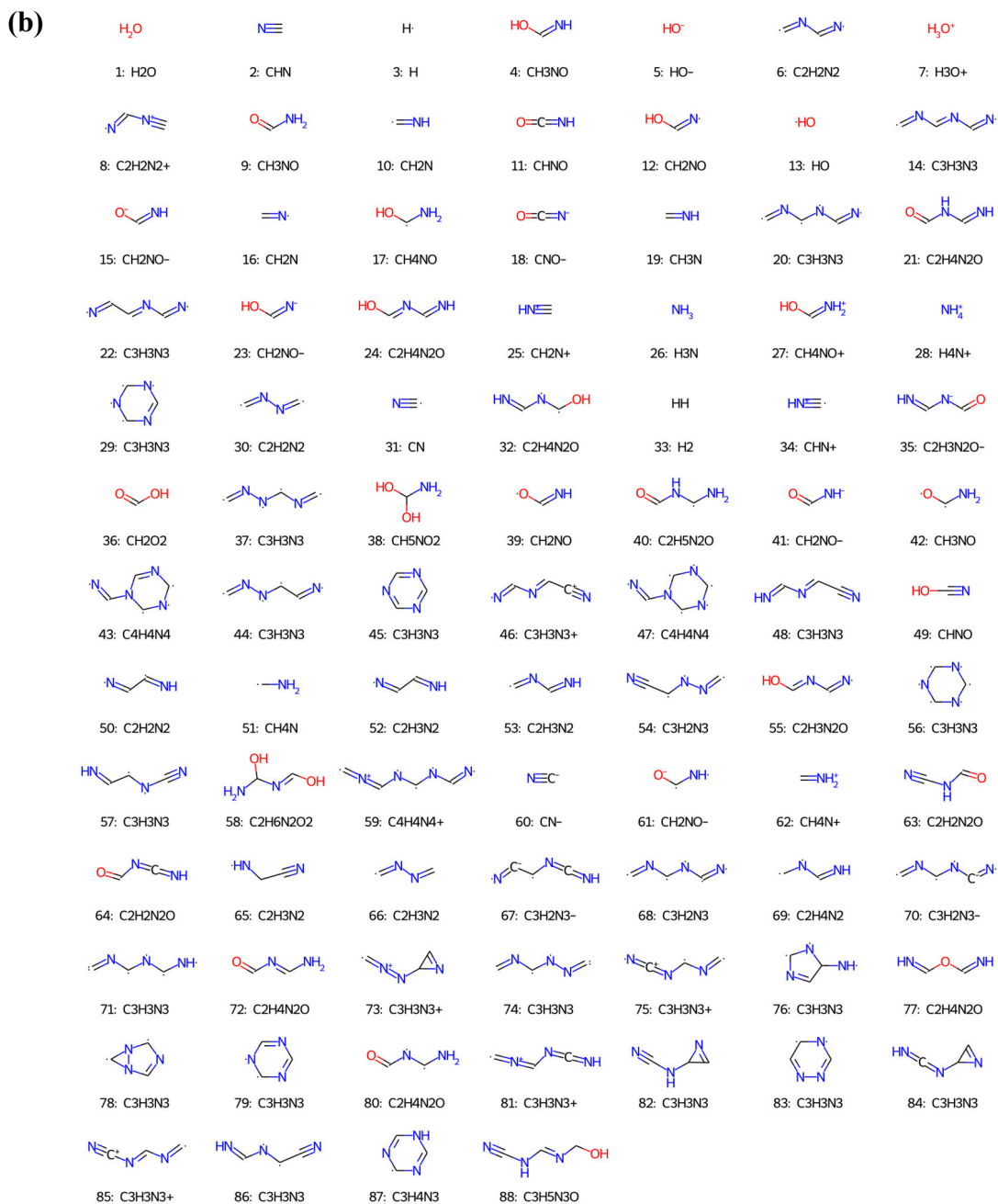

**Figure 4.** The chemical transformation network (a) automatically formed based on reaction events, excluding the reactions occurring less than three times in the MTD. Each node represents a species. Their sizes and colors are based on the times of their participation in the chemical transformations. Their index corresponding structures and chemical formula are shown in (b). Edges encode chemical transformations, and their colors and widths are set based on their corresponding transformation times.

As expected, the overall network is extremely complex, and dozens of species are involved. A few of important reaction events occur many times in the MTD simulations. For instance, hydrogen ions (or hydrogen radicals) and hydroxyl ions (or hydroxyl radicals) form in large quantities due to the abundance of $H_2O$ molecules in the system. The reaction events between $H_2O$ molecules and HCN molecules yield various species such as alcohols, aldehydes, and lipids. At the same time, the HCN polymerization reactions are observed to give HCN dimers and trimers. Many of these above major reaction events involve the small-sized or medium-sized compounds, which take place often in the early MTD simulations. As the contrast, the large compounds produced by the polymerization often appear in the later stage of the MTD simulations. As they often involve different species, and we cannot assign them as the "same" reaction. Due to our selection rule to plot the chemical transformation network, the polymerization reactions and its large-sized products are missing in Figure 4. In another word, the current chemical transformation network tends to give reactions with small-sized and medium-sized species involved in the early stage of dynamics. The complete list of all detected species is given in the Supporting Information.

To provide more transparent overview on the chemical transformation of all species, we listed some critical species in Figure 5. We mainly derived the involved species to five types according to their structural features and formation sequence.

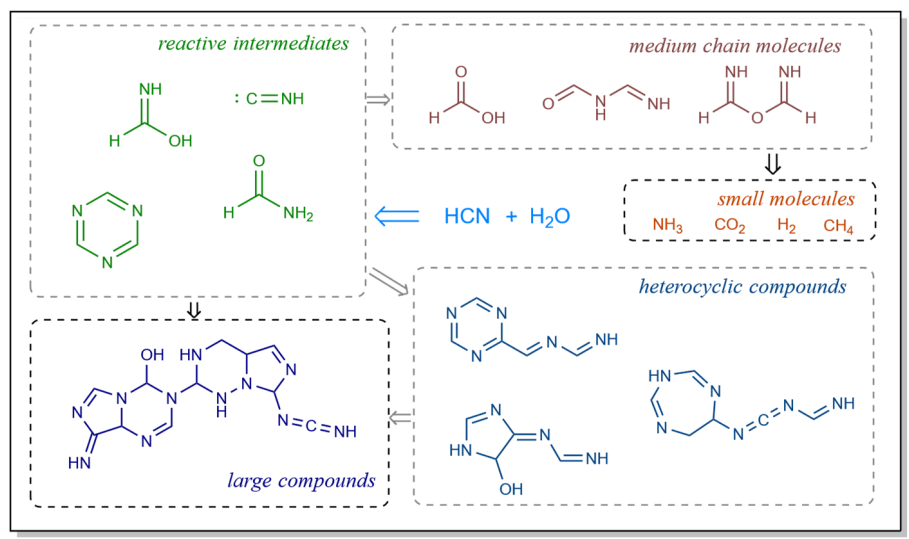

**Figure 5.** Overview of species involved in the HCN and H$_2$O reactions.

Starting from the fundamental reactants, various reactive intermediates are produced, such as formimidic acid (NH=CHOH), formamide (NH$_2$CH=O) and 1,3,5-triazine (C$_3$H$_3$N$_3$). These intermediates, in turn, participate in further reactions, leading to the generation of the medium chain molecules like formic acid (HCOOH) or heterocyclic compounds. Then the dissociation reactions of medium chain molecules could form small molecules such as CO$_2$, NH$_3$, and H$_2$. Moreover, large compounds with C-N chains and O-contained rings yield in the addition and isomeric reactions of heterocyclic compounds.

After obtaining the overview of the chemical transformation, we further sought for the detailed understanding of major important pathways in the reactions between HCN and H$_2$O. First, we focused on the major pathways that lead to the formation of the medium-chain molecules, as shown in Figure 6. We found two sets of primary reactions:

one initiated by the interaction between H₂O and a single HCN molecule, and the other initiated by the interaction between H₂O and two HCN molecules.

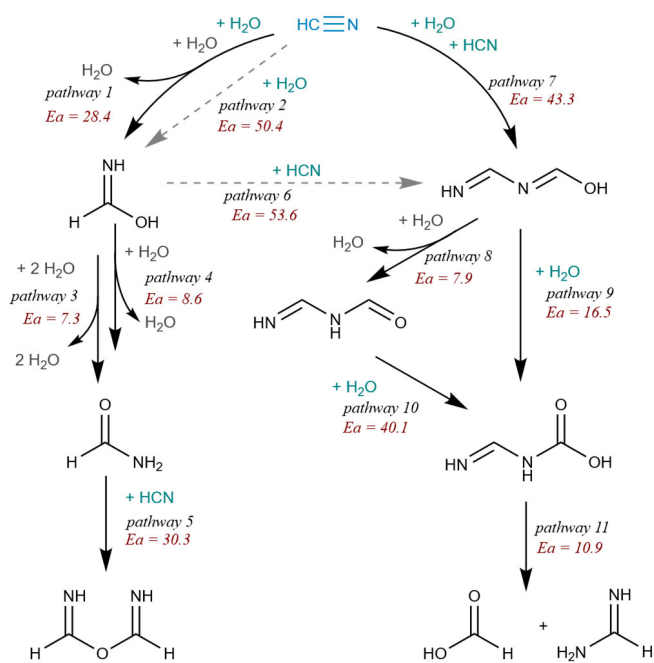

**Figure 6.** The reaction mechanism network starting from fundamental reactants HCN and H₂O (highlighted in blue). Each elementary step is labeled by its index number, and its activation energy (*Ea*) is given in the unit of kcal/mol. The dashed arrow line indicates a barrier higher than 50 kcal/mol.

In the first set of reactions, two pathways (Pathway 1, 2) can form the species NH=CHOH. Pathway 1, facilitated by one H₂O serving as the proton shuttle, exhibits a significantly lower energy barrier compared to Pathway 2. Similarly, in Pathway 3 and Pathway 4, either one or two molecules of H₂O serve as the proton shuttles, effectively facilitating the transformation of NH=CHOH to its tautomer NH₂CH=O, with a barrier energy below 10 kcal/mol. NH$_2$CH=O can then react with HCN through Pathway 5 of the C-O addition, which is characterized by a barrier of 30.3 kcal/mol.

In the second set of reactions, the species NH=HC-N=CHOH is formed through the addition of two HCN molecules and one H$_2$O via Pathway 7 with the 43.5 kcal/mol barrier. For comparison, the NH=CHOH and HCN addition reaction via Pathway 6 is less favorable due to a higher barrier (53.6 kcal/mol). With the assistance of H$_2$O via Pathway 8, the compound NH=HC-N=CHOH transforms into its isomer NH=HC-NH-CH=O. Then an unstable species C$_2$H$_6$N$_2$O$_2$ can form through Pathway 9 or Pathway 10. Finally, this species may undergo the decomposition through Pathway 11 with a barrier of 10.9 kcal/mol, leading to the formation of HCOOH and formimidamide (NH$_2$-CH=NH).

Additionally, the dimer and trimer formation reactions of HCN molecules which served as the crucial step to yield heterocyclic compounds were important reactions. We investigated the relevant mechanisms and found multiple pathways. Here, we constructed the reaction mechanism network of them in Figure 7, which mainly displays the pathways with the reaction barriers less than 56 kcal/mol.

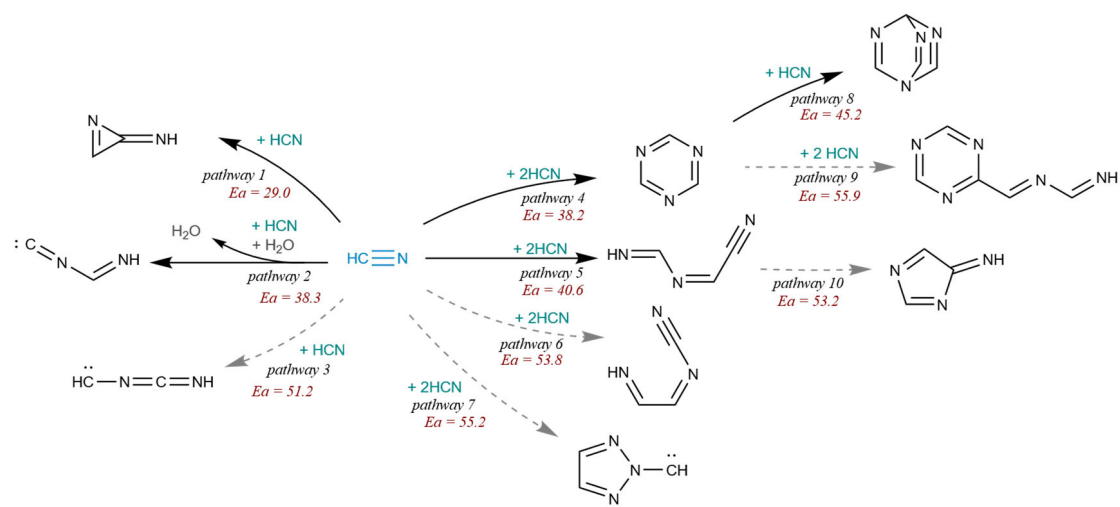

**Figure 7.** The reaction mechanism network starting from fundamental reactants HCN (highlighted in blue). Each elementary step is labeled by its index number, and its activation energy (*Ea*) is given in the unit of kcal/mol. The dashed arrow line indicates a barrier higher than 50 kcal/mol.

For the HCN dimer formation reactions, Pathway 1 is the lowest barrier with 29.0 kcal/mol. Additionally, aided by $H_2O$ as the proton shuttle, HCN molecules can form a chain-type dimer via Pathway 2 with a barrier of 38.3 kcal/mol. Because above pathways are endothermic reactions, the dimers may decompose back to HCN molecules again. As these dimers are not very stable, the HCN trimers should be more important species in the formation of the large compounds.

For the HCN trimer formation reactions, Pathway 4 displays the lowest barrier of 38.2 kcal/mol, which gives a hexabasic hexacyclic species, $C_3H_3N_3$. With the further additions with HCN (Pathway 8, 9), this species could generate larger species, including an adenine isomer via Pathway 9. Additionally, Pathway 5 with a barrier of 40.6 kcal/mol leads to the formation of the C-N chained-type species that can further yields a pentacyclic species via the cyclization reaction.

Expect the above reactions, the formation of the large compounds was also widely observed in the MTD simulation. We presented a typical example trajectory in Figure 8. In the early phase of the simulation, most species consist of 3 or 6 atoms, as the initial system is composed of HCN and $H_2O$ molecules. With the trajectory propagation, the size of molecules in the system steadily increases, while the number of small molecules with 3 atoms ($H_2O$ or HCN) decrease. After the 5 ps evolution, compounds containing 9 atoms begin to form, due to the HCN trimer formation. As the simulation continue, a single large species, comprising more than 20 atoms and even approaching 40 atoms, are observed. This phenomenon may account for many HCN molecules with the triple bonds in the current system. As the total electron is conserved for the whole system, it is very easy for these HCN molecules to form the longer chain or ring structures.

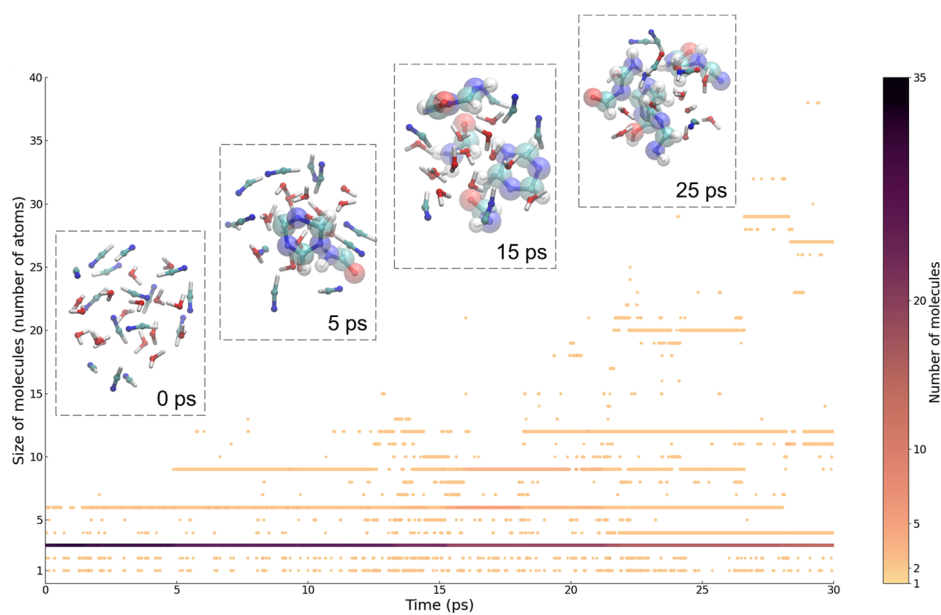

**Figure 8.** Timeline of molecular sizes represented by the number of atoms and the number of molecules. The timeline obtains from one typical trajectory for the reactions between HCN and $H_2O$ molecules. The insets show the snapshots at the given time point.

Our current approach generated the reaction network that was comparable with the previous studies, while some difference also existed. For instance, some additional mechanisms of the dimer formation of HCN molecules were found in our current work comparing with the previous work by Anoop et.al.[31] While, some channels were not the primary ones in our calculations comparing the previous high-temperature and high-pressure nanoreactor work by Vanka et.al.[84] Therefore, these differences should be paid attention.

We attributed such different findings to the ways to run the dynamics. The high-temperature and high-pressure nanoreactor method applies a piston potential that compresses molecules towards the center, giving the very strong molecular interaction and very high system energy to induce the cleavage of the chemical bonds. Therefore, this method tends to emphasize the formation of small molecules. However, our employed MTD method applies the biasing potential, which primarily drives the system away from its referred structures. This seems to result in polymerization of large chain or ring structures. Therefore, different products may come out from these two methods. As the exploration of chemical space is a N-P hard problem, both two methods should be complimentary to each other to achieve the complicated investigation. In addition, more discussions and analyses on their different outcomings and the underline physical insight should be an interesting research topic in the future.

## 5. CONCLUSION

In this work, we proposed an approach for the automated exploration of the reaction mechanism and network for complex multi-species systems. This approach demonstrates its great applicability and effectiveness. The employment of the MTD (RMSD) / GFN2-xTB method can efficiently explore the reaction space of the given system. The reaction events are extracted by considering the molecular connectivity, which provides good initial guess to conduct the further **TS** searches at more accurate electronic structure levels. Additionally, the utilization of HMM largely reduces the computational cost in the MEP searches. Finally, two different types of reaction networks are established, which are based on the reaction events found in trajectory propagation and the detailed reaction mechanism with MEPs, respectively.

In this work, we tested the feasibility of our approach by its performance in two typical reaction systems, *i.e.,* bimolecular reactions between HCHO and $NH_3$ and multi-species reactions between HCN and $H_2O$. The reaction mechanisms were explained, and their reaction networks were also constructed. In former system, we found the primary pathways similar with previous studies. In later one, the chemical transformation network displayed the complex transformation among a huge number of species. We also investigated the detailed mechanisms of several typical reactions, such as the production of medium-chain molecules, the HCN oligomerization, and the formation of large compounds. Overall, the current automated approach provides a rather practical approach to study the complex reaction network.

Our long-term goal is to develop the automated approach for the exploration of the chemical reaction network for complex systems, which can largely reduce human trial. To realize such task, some challenging problems need to be solved in the future. The current sampling method is suitable for systems with low-reactivity, but not for systems with high-reactivity. For instance, when the system is composed of many free radicals, these species may start to react to each other even in the sampling process. Meanwhile, the treatment of the reactions involving the species with higher multiplicity is even more challenging. In addition, the suitable choice of the MTD setting is still not a fully trivial task. We can also apply other enhanced sampling methods[103-105] to generate reaction events. To further reduce computational cost, it is preferable to employ machine learning potential energy surface[106-109] in reactive MD. All of them are very challenging topics in the future. With the further development and improvements, we believed our approach could become a highly automatic and user-friendly tool in the exploration and discovery of reaction mechanism for complicated multi-species systems.

## ASSOCIATED CONTENT

**Supporting Information**

Several relevant information: Identification of reactants and products in a reaction event; Additional discussions on reaction event identification; Initial samplings; Test calculations of MTD; MTD simulations and Identified species in HCHO and $NH_3$ reactions and HCN and $H_2O$ reactions.

## Author Information

**Corresponding Author**

E-mail: zhenggang.lan@m.scnu.edu.cn; zhenggang.lan@gmail.com.

**Notes**

The authors declare no competing financial interest.

All codes are variable by contacting with the corresponding authors.

## Acknowledge

This work was supported by NSFC projects (No. 21933011 and 21873112) and the Opening Project of Key Laboratory of Optoelectronic Chemical Materials and Devices of Ministry of Education, Jianghan University (JDGD-202216). The authors thank the Supercomputing Center, Computer Network Information Center, Chinese Academy of

# Supporting Information for

# Automated Exploration of Reaction Network and Mechanism via Meta-dynamics Nanoreactor

*Yutai Zhang[1], Chao Xu[1], Zhenggang Lan[1\*]*

[1] Guangdong Provincial Key Laboratory of Chemical Pollution and Environmental Safety and MOE Key Laboratory of Environmental Theoretical Chemistry, SCNU Environmental Research Institute, School of Environment, South China Normal University, Guangzhou 510006, P. R. China.

\* Corresponding Author

E-mail: zhenggang.lan@m.scnu.edu.cn; zhenggang.lan@gmail.com

# 1. Details of MTD (RMSD) method

The meta-dynamics (MTD) was performed at the GFN2-xTB level in xtb software package.[1] As suggested by Grimme,[2] the root-mean-square deviation (RMSD) is served as collective variable (CV) labelled as $\Delta$, and its definition is given as:

$$\Delta_i = \sqrt{\frac{1}{N}\sum_{j=1}^{N}\left(r_j - r_j^{ref,i}\right)^2}$$

where $r_j$ is the Cartesian coordinates of the atom in current time step, $r_j^{ref,i}$ is the corresponding coordinates in reference structure $i$, and $N$ is the number of atoms. With comparison with the reference structures, RMSD-CV is calculated, and the biasing potential is added along this CV. The biasing potential is given as:

$$E_{bias}^{RMSD} = \sum_{i=1}^{n} k_i exp(-\alpha_i \Delta_i^2)$$

where $n$ is the number of reference structures, $k_i$ is the strength parameter and $\alpha_i$ is the parameter that determines the width of the biasing potential.

## 2. Wall potential in xtb software package

In the molecular dynamics (MD) and MTD simulations, the wall potential implemented in the xtb software package was employed to constrain the molecules not escaping from the reactor. The wall potential is logfermi format and its definition is given as follows:[3]

$$V = \sum_A k_B T \log\{1 + \exp[\beta(|R_A - O| - R_{sphere})]\}$$

where $k_B$ is the Boltzmann constant, $T$ is formally the temperature for scaling the strength of the potential and its default value is 6000 K, $\beta$ is the steepness of the potential and its default value is 10 Bohr$^{-1}$, $R_A$ are the Cartesian coordinates of atom A, $O$ is the center of the reactor and $R_{sphere}$ is the radius of the sphere used for confining. In this work, we mainly adjusted $\beta$ to change the softness of wall potential and modified $R_{sphere}$ to define the size of reactor.

## 3. Identification of reactants and products in a reaction event

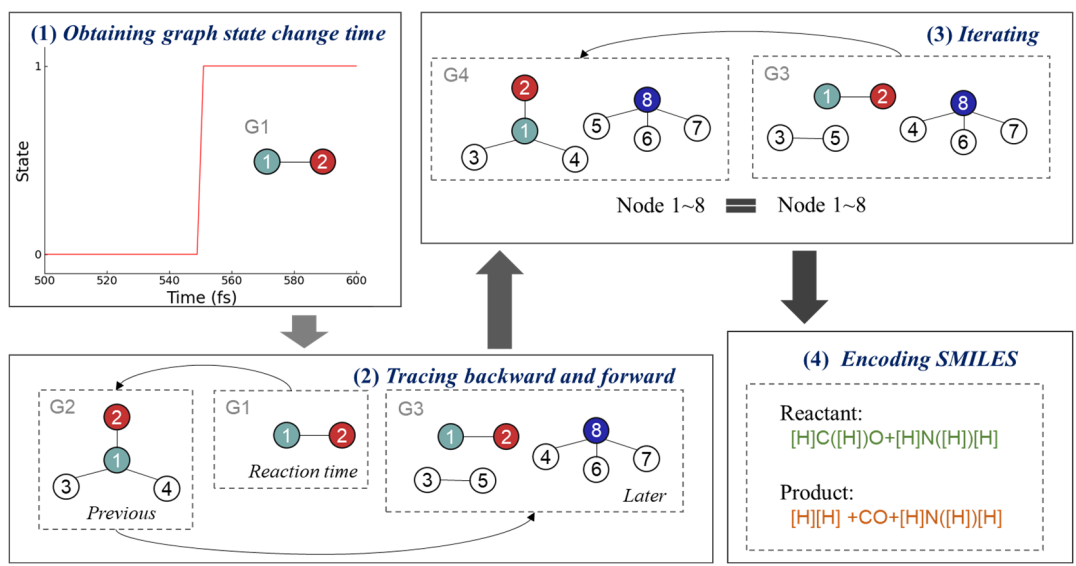

**Figure S1.** Identification scheme of the reactants and products in a reaction event follows several steps: (1) obtaining the time of the graph-state change, (2) tracing backward and forward, (3) iterating (4) encoding SMILES. A HCHO + NH$_3$ → H$_2$ + CO + NH$_3$ reaction event serves as an example, where red cycle is O atom, pale green cycle is C atom, white cycle is H atom, and deep blue cycle is N atom.

Here we take a HCHO + NH$_3$ → H$_2$ + CO + NH$_3$ reaction event as an example to illustrate how to identify the reactants and products, as shown in Figure S1.

(1) After the filter by the HMM, the signal transition of 0→1 means a reaction event take place. This step also provides a new-born subgraph G1, which corresponds to the fact that a new molecule (CO) is generated by the current reaction event as a product.

(2) Starting from the current transition time, we try to get all species involved in

the current reaction event. We trace all involved nodes 1, 2 (corresponding to atom C, O, found in Step 1 respectively) backward along the time-axis to find which subgraphs these nodes belong to. If these nodes are the part of the subgraph $G_2$ (HCHO) at the earlier time step, it means that more atoms participate in this event. We included all nodes of the newfound subgraph to the define the new involved atoms (node 1-4 corresponding to C, O, H and H, respectively). Up to now, all four atoms in HCHO were included, as shown in Figure S1.

(3) We trace forward along the time-axis and try to identify which subgraphs cover these involved nodes. In the given example, we tried to find which molecules contain four atoms (H, C, H and O) identified in Step 2. Now three compounds are found, *i.e.,* one including nodes 1 and 2, one including node 3 and node 5, and one including node 4, 6, 7 and 8. Now, all atoms in these three compounds are included to start next tracing.

(4) After iterating the above forwards and backwards procedure several times until no new node is found, all involved subgraphs are found, which define all reactants and products. In the given examples, reactants are HCHO and $NH_3$, while products are $H_2$, CO and $NH_3$.

(5) We transform all involved reactant and product graphs into canonical SMILES (simplified molecular input line entry specification) code by employing RDKit[4] Python module.

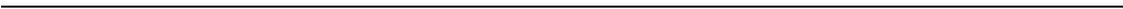

# 4. Additional discussions on reaction event identification

In the current work, the effectiveness of our reaction event identification method relies on a set of parameters. Here, we provide a brief discussion on these parameters.

(1) **Dump time step in MTD:** Currently, we applied time steps of 2 fs and 5 fs to save the molecular structures in the MTD run. The selection of dump time step affects the identification of reaction events. Generally, the larger dump time step leads to the less identification of reaction events. The dump time step smaller than 10 fs is recommended.

(2) **Molecular connectivity:** Our current method utilized the standard in OpenBabel[5] Python module to judge whether a bond exists or not for pairs of atoms. The similar approach was also applied in previous study.[6] Some other works employed different criteria, such as a certain coefficient multiplying with bond lengths[7] or atomic covalent radii.[8] The differences between these standards are usually not significant.

(3) **HMM parameters**: T (transition probability) and O (emission probability) in hidden Markov model (HMM) play a role in filtering the reaction events. Smaller values in the off-diagonal of T and larger values in the off-diagonal of O enhance the filter effect.

(4) **Tracing time range for reaction events**: Currently, our tracing time range included a total of 60 snapshots from trajectory, with 30 snapshots taken forward and 30 snapshots taken backward. Selecting the too large time range may lead to that fact that several different sets of reactants and products are found for the same event. In addition, the selection of the time range should be compatible with the MTD dump time.

## 5. Transition State Searches

The current transition state (**TS**) search strategy comes from our previous method.[7] In the current work, two settings may influence the successful possibility of **TS** searches.

(1) The first one is the number of initial guess structures that are directly taken from the trajectory. Increasing the number of guess structures enhances the likelihood of finding **TS**, but it comes with a higher computational cost.

(2) The second one is the interval of selecting guess structures. In our work, the guess structures extracted from snapshots near the reaction time show a higher chance to find the **TS** structures.

We also applied two double-ended **TS** search methods to help locate **TS**: the climbing image nudged elastic band (CI-NEB) method[9] in ORCA software package[10] and the pushing/pulling RMSD bias potentials (RMSD-PP)[2] method in xtb software package. In this work, these two approaches seemed not work as well as the above direct approach. When the situations became more complicated, we did not know which approach is better. In this case, the trial calculations became useful to select the proper **TS** search methods.

## 6. Initial samplings

We applied the MD method to obtain the initial structures for MTD. In the MD run, some common settings were adopted: the equilibrium temperature was set at 300 K, the MD time step were 1 fs, and the dump time step of saving trajectory was 500 fs.

For HCHO and $NH_3$ bimolecular reaction, we first constructed a sphere with a radius of 4.3 Å containing one HCHO molecule and one $NH_3$ molecule, and the whole system was optimized at the GFN2-xTB level. Next, we performed the MD at the GFN2-xTB level for 500 ps starting from the optimized structures, combined with the wall potential of the same 4.3 Å radius. Here we chose the geometries from the MD after equilibrium to define the initial configurations for the next MTD steps.

For HCN and $H_2O$ muti-species reaction, we constructed three different spheres with radius of 7.8 Å or 3.5 Å totally containing 16 HCN molecules and 15 $H_2O$ molecules. The system settings were following the manners of the high-temperature and high-pressure nanoreactor work by Vanka et al,[11] which employed the piston potential with 3.5~10 Å radius. After the optimization of these systems, three sets of MD simulations were run for 30 ps with the wall potential. All initial configurations were obtained from the trajectories after equilibrium.

## 7. Test calculations of MTD

In MTD test calculations, some common settings were employed:

(1) $\alpha_i$ was set to 0.7 Bohr$^{-2}$,

(2) all atoms of entire system were considered in the RMSD-CV calculations,

(3) we only selected one reference structure that was updated every 1 ps.

### 7.1 Reactions between HCHO and NH$_3$

For reactions between HCHO and NH$_3$, we only tested the $k_i$ parameter. All details are given in Table S1.

**Table S1.** Employed settings for test MTD simulations in HCHO and NH$_3$ reactions. The rate of reactive MTD trajectories ($R_{\text{reactive trajectory}}$) in each setting is shown.

| HCHO and NH$_3$ Reactions (50 trajectories for 10 ps) | | | | |
|---|---|---|---|---|
| *Label* | 1 | 2 | 3 | 4 |
| $k_i$ / $E_h$ | 0.10 | 0.15 | 0.20 | 0.25 |
| $R_{\text{reactive trajectory}}$ | 0% | 0% | 12% | 76% |

We briefly discussed the results as follows:

(1) When $k_i$ was set to 0.1 and 0.15 $E_h$, no reaction event was observed.

(2) When $k_i$ was set to 0.2 $E_h$, some reactions were observed. When we further raised this value to 0.25 $E_h$, most trajectories show reactions.

Therefore, these two values (0.2 $E_h$, 0.25 $E_h$) were chosen to continue the extensive MTD work.

**7.2 Reactions between HCN and H$_2$O**

For reactions between dozens of HCN and H$_2$O molecules, our tests not only involved the $k_i$ parameter but also other parameters, including the bath temperature ($T_{bath}$) and the radius of the wall potential ($R_{sphere}$). All details are given in Table S2.

**Table S2.** Employed settings for test MTD simulations in HCN and H$_2$O reactions.

| HCN and H$_2$O Reaction (10 trajectories for 30 ps) | | | | |
|---|---|---|---|---|
| *Label* | 1 | 2 | 3 | 4 |
| | | | | *MTD part* |
| $k_i$ / $E_h$ | 0.4 | 0.4 | 0.4 | 0.8 |
| $\Delta t$ / fs | 0.5 | 0.5 | 0.2 | 0.2 |
| $T_{bath}$ / K | 300 | 300 | 1500 | 1500 |
| | | | | *Wall potential part* |
| $\beta$ / Bohr$^{-1}$ | 10 | 0.5 | 0.5 | 0.5 |
| $R_{sphere}$ / Å | 7.8 | 3.5 | 3.5 | 3.5 |

The test results are briefly described as follows:

(1) Initially, we set $T_{bath}$ to 300 K and $R_{sphere}$ to 7.8 Å. In this case, the reactive probability was very low.

(2) We decreased the sphere radius to 3.5 Å to increase reactivity and set $\beta$ to 0.5 Bohr$^{-1}$. This $\beta$ setting was necessary to maintain the stability of the trajectory propagation in the MTD. Under these conditions, we observed that the HCN polymerization was dominant, and most reactions did not include $H_2O$.

(3) We increased the $T_{bath}$ to 1500 K. To maintain the MTD stability, we adjusted the MTD time step to 0.2 fs. Under such setting, more reactions between $H_2O$ and HCN occurred, while the HCN polymerization was still predominant.

(4) We increased the $k_i$ parameter to 0.8 $E_h$. However, in this setting, the MTD often collapsed during the trajectory propagation, and we thought that $k_i$ should not be raised further. However, we observed that some small molecules generated under these conditions.

Consequently, we decided to use $k_i$ = 0.6 $E_h$ and 0.8 $E_h$ to get all types of reactions, including reactions between $H_2O$ and HCN, the HCN polymerization and the generation of small molecules. The test calculation also indicated that the low-temperature and the small biasing potential tended to induce the HCN polymerization. Increasing the temperature and biasing potential leaded to more reactions between $H_2O$ and HCN. When the reaction condition became more radical, small compounds were generated.

## 8. MTD simulations

In the current MTD simulations of both systems, two $k_i$ values were taken, which correspond to the small and large biasing potentials, respectively. With the same setting, we first ran a set of trajectories to explore the reactions, and then ran the second set of trajectories to check the convergence. The trajectory groups used for simulation and validation test for high and low biasing potentials were labelled as $H_1$, $H_2$, $L_1$, $L_2$, respectively.

**8.1 Reaction between HCHO and NH$_3$**

For HCHO and NH$_3$ bimolecular reaction, the MTD setting is show in Tabel S3. In our test calculations, the system reactions became violent with $k_i = 0.25$ $E_h$, so we applied the short MTD simulation time of 5 ps. With $k_i = 0.2$ $E_h$, the reactivity was rather weak, a longer MTD simulation time of 10 ps was used.

Taking the high biasing potential as an illustrative example, the major reaction events in the Group $H_2$ were found in the Group $H_1$. Meanwhile, the major reaction events in the high biasing potential Group ($H_1$ and $H_2$) basically contained all major reaction events in low biasing potential Group ($L_1$ and $L_2$). Therefore, we believed that the MTD calculations of this system were converged.

**Table S3.** Parameter settings in the MTD simulations of HCHO and NH$_3$ reaction.

| | HCHO and NH$_3$ Reaction | | | |
|---|---|---|---|---|
| *Label* | *H$_1$* | *H$_2$* | *L$_1$* | *L$_2$* |
| $N_{trajectory}$ | 500 | 100 | 500 | 100 |
| $N_{molecule}$ / (HCHO : NH$_3$) | 1:1 | 1:1 | 1:1 | 1:1 |
| | | | | ***MTD part*** |
| $k_i$ / $E_h$ | 0.25 | 0.25 | 0.2 | 0.2 |
| $α_i$ / Bohr$^{-2}$ | 0.7 | 0.7 | 0.7 | 0.7 |
| $Δt$ / fs | 0.5 | 0.5 | 0.5 | 0.5 |
| $t_{total}$ / ps | 5 | 5 | 10 | 10 |
| $T_{bath}$ / K | 300 | 300 | 300 | 300 |
| | | | | ***Wall potential part*** |
| $β$ / Bohr$^{-1}$ | 10 | 10 | 10 | 10 |
| $R_{sphere}$ / Å | 4.3 | 4.3 | 4.3 | 4.3 |

## 8.2 Reaction between HCN and H₂O

For the HCN and H2O muti-species reaction, the employed setting is show in Table S3. In our test calculations, we found most reaction events happening within 20 ps and the total simulation time was set to 30 ps. Due to the instantaneous high temperature and the convergence failure in the electronic structure calculations, many simulations were not completed. Therefore, we employed the available parts of the trajectories for analysis.

- For $k_i$ = 0.8 $E_h$, 30 trajectories were analyzed, where 15 trajectories had a length of 30 ps and 15 trajectories had a length of 20 ps. 5 additional trajectories of 30 ps were analyzed for validation test.

- For $k_i$ = 0.6 $E_h$, 30 trajectories were first analyzed, where 15 trajectories had a length of 30 ps, and 15 trajectories had a length of 20 ps. 5 additional trajectories of 30 ps were analyzed for validation test.

**Table S4.** Parameter settings in the MTD simulations of HCN and H$_2$O reaction.

| | HCN and H$_2$O Reaction | | | |
|---|---|---|---|---|
| *Label* | H$_1$ | H$_2$ | L$_1$ | L$_2$ |
| $N_{trajectory}$ | 15/15 | 5 | 15/15 | 5 |
| $N_{molecule}$ / (HCN : H$_2$O) | 15:16 | 15:16 | 15:16 | 15:16 |
| | | | | ***MTD part*** |
| $k_i$ / $E_h$ | 0.8 | 0.8 | 0.6 | 0.6 |
| $\alpha_i$ / Bohr$^{-2}$ | 0.7 | 0.7 | 0.7 | 0.7 |
| $t_{total}$ / ps | 30/20 | 30 | 30/20 | 30 |
| $\Delta t$ / fs | 0.2 | 0.2 | 0.2 | 0.2 |
| $T_{bath}$ / K | 1500 | 1500 | 1500 | 1500 |
| | | | | ***Wall potential part*** |
| $\beta$ / Bohr$^{-1}$ | 0.5 | 0.5 | 0.5 | 0.5 |
| $R_{sphere}$ / Å | 3.5 | 3.5 | 3.5 | 3.5 |

For both parameter settings in the MTD simulations, the following rule was used to judge the convergence.

(1) One of important reactions was the HCN polymerization, leading to the formation of large compounds. Although important, many of polymerization only took place once or twice in the MTD, if we judged the frequency of these reactions by monitoring each single compound. In this case, we should perform two independent analyses: *i.e.,* one focusing on the polymerizations and one focusing on other reactions.

(2) For convergence test, we only focused on the second types of reactions without

considering the polymerization. In this case, we defined other reactions involving the small compounds as major reactions, if they occurred more than twice.

(3) The major reaction events found in Group $H_2$ and $L_2$, were already discovered in Group $L_1$ and $H_1$, respectively, which indicated that more trajectories were needed in the MTD with the current parameter setting.

(4) Comparing two sets of MTD, the major reaction events appearing in the $H$ group contained the ones in the $L$ group. We assumed that the MTD calculations for this system were converged.

## 9. The chemical transformation network of HCHO and NH₃ reactions

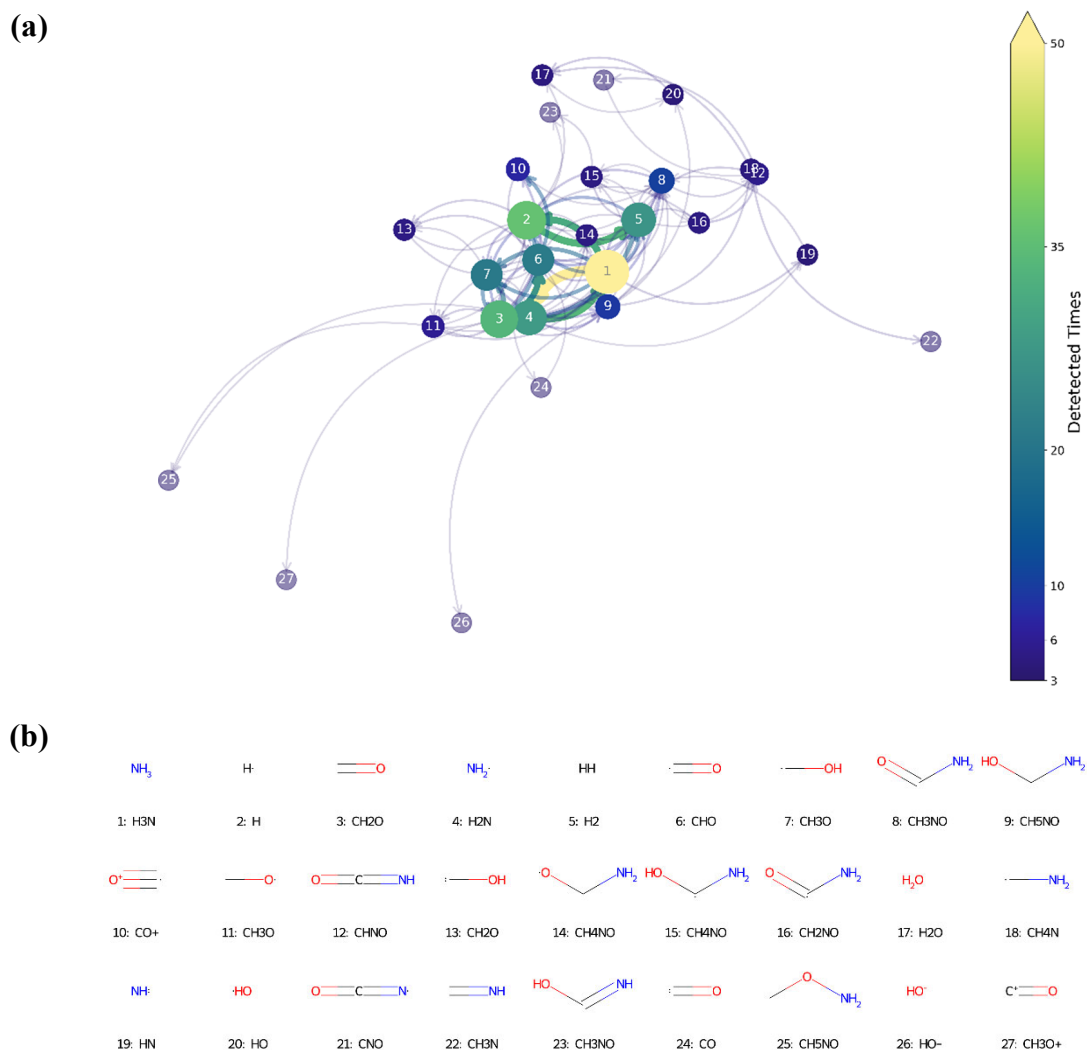

**Figure S2.** The chemical transformation network (a) automatically formed based on reaction events of HCHO and NH₃ reactions, excluding the reactions occurring less than three times in the MTD. Each node represents a species. Their sizes and colors are based on the times of their participation in the chemical transformations. Their index corresponding structures and chemical formula are shown in (b). Edges encode chemical transformations, and their colors and widths are set based on their transformation times.

## 10. HCHO and NH₃ primary reactions

**Figure S3.** TS structures and product structures involved in the HCHO and NH₃ primary reactions. The index numbers of these structures correspond to the reaction indices given in Figure 3.

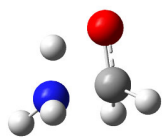

TS1

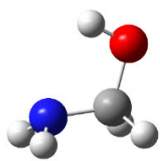

P1

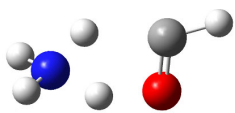

TS2

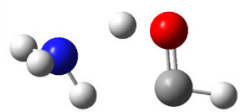

P2

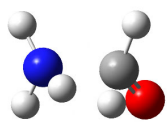

TS3

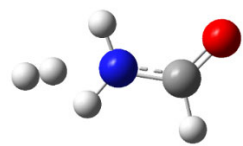

P3

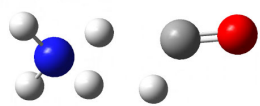

TS4

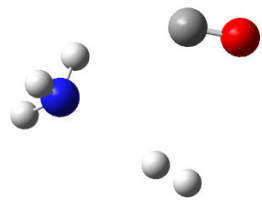

P4

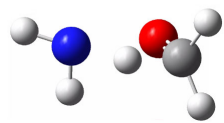

TS5

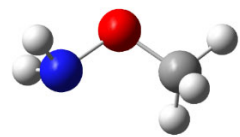

P5

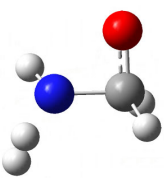

TS6

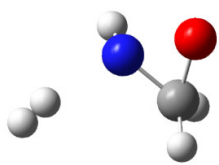

P6

## 11. Detected species in HCHO and NH₃ reactions

**Figure S4.** All involved species identified in the HCHO and NH$_3$ reactions. The structure and the chemical formula of each species is given as follows. All index numbers of species are sorted according to the frequency of their participation in the reaction events in the MTD dynamics, with smaller index numbers indicating higher participation frequency.

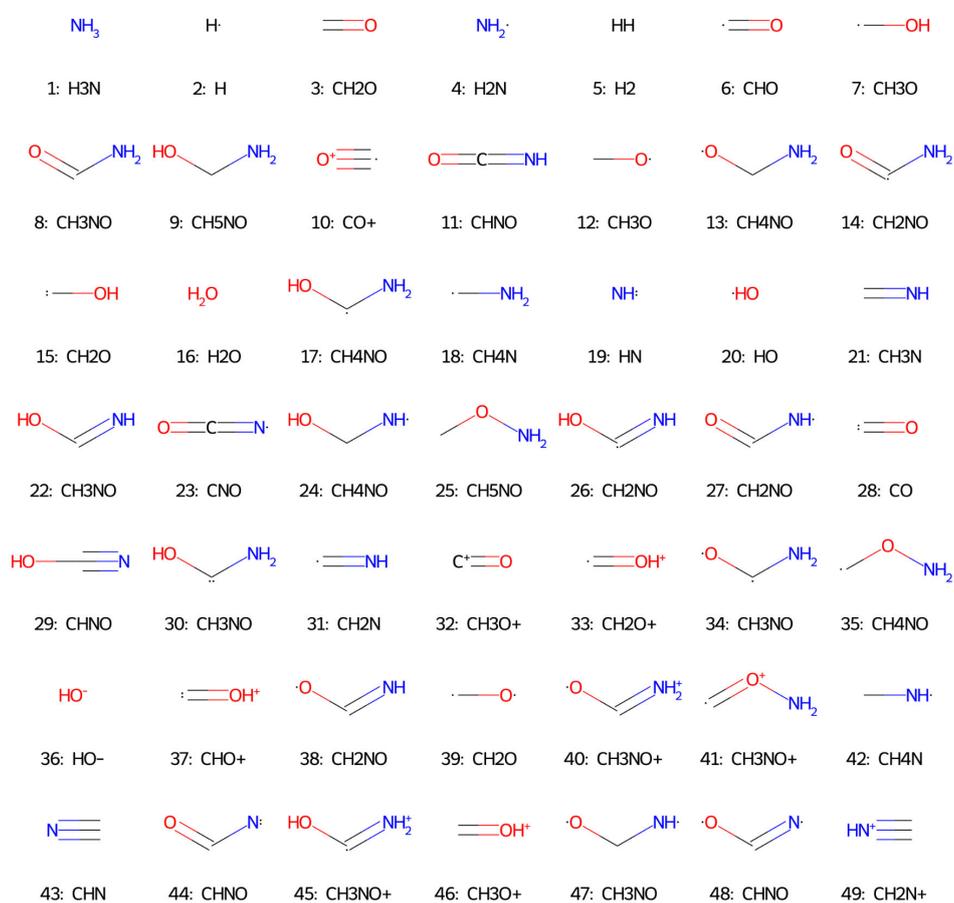

## 12. Detected species in HCN and H₂O reactions

Because a huge number of species were found in HCN and H$_2$O reaction network, we list them in grid pictures format in the second supporting file. All of them were sorted according to their molecular size and the frequency of their participation in reactions in the MTD.